\documentclass[]{aa}

\usepackage{graphicx}

\usepackage{txfonts}
\begin{document}

   \title{A self-consistent and time-dependent hybrid blazar emission model}

   \subtitle{Properties and application}

   \author{M. Weidinger
          \inst{1}
          \and
          F. Spanier\inst{2}
          }

   \institute{Lehrstuhl f\"ur Theoretische Physik IV, Ruhr-Universit\"at Bochum, Universit\"atsstr. 150, 44780 Bochum\\
              \email{mweidinger@tp4.rub.de}
         \and
             Centre for Space Research, Private Bag X2600, Potchefstroom Campus, North-West University, Potchefstroom 2520\\
             \email{felix@fspanier.de}
             }

   \date{Received March 15, 2014; revised: October 8, 2014; accepted: October 9, 2014}


  \abstract
   {}
   {A time-dependent emission model for blazar jets, taking acceleration due to Fermi-I and Fermi-II processes for electrons and protons
   as well as all relevant radiative processes self-consistently into account, is presented.}
   {The presence of highly relativistic protons
   within the jet extends the simple synchrotron self-Compton case not only in the very high energy radiation of blazars, but also in the
   X-ray regime, introducing non-linear behaviour in the emitting region of the model by photon-meson production and emerging electron positron
   pair cascades.}
   {We are able to investigate the variability patterns of blazars in terms of our model in all energy bands, thus narrowing down the parameters used.
   The blazar \mbox{1 ES 1011+496} serves as an example of how this model is applied to high frequency peaked BL Lac Objects in the presence of non-thermal protons within the jet. Typical multiband patterns are derived, which are experimentally accessible.}
   {}

   \keywords{acceleration of particles --- galaxies: active --- galaxies: jets --- quasars: general --- quasars: individual(1ES1011+496) ---
   gamma rays: general
               }
   \maketitle
%

\section{Introduction}
The origin of the characteristic highly non-thermal radiation from blazars, a subclass of active galactic nuclei (AGN) with the highly relativistic outflow emerging under a small angle to the line of sight, is undoubtedly the jet. The two prominent humps in the spectral energy distribution (SED) of these objects, one in the optical to X-ray regime and a second one occurring at the highest energies up to TeV, require high beaming factors which only the jet can provide \citep{urry01}. Blazars have gained a lot of interest over the past decades mainly as a result of the discovery of their variable very high energy (VHE) emission with variations of many orders of magnitudes in the TeV regime using Air-Cherenkov telescopes like H.E.S.S., MAGIC, and VERITAS entrenching gamma-ray astronomy as its own field of research, from \citet{punch} to \citet{aha}. With the Fermi satellite and its all-sky survey capabilities (in orbit since 2009), the availability of multiwavelength (MWL) data (combined with radio, optical, and X-ray
observations using the Swift satellite, RXTE, Chandra etc.) has rapidly increased so that the variable emission of AGN can be monitored in different energy bands. This data and the steadily increasing number of discovered blazars provides a solid basis for systematic investigations with theoretical emission models.\\
There have been many attempts to classify blazars, for example using visible line emission \citep{strittmatter, marcha, urry01}, which in the end was summarized in the blazar sequence, first proposed by \citet{fossati01}. Although there are exceptions, the most luminous blazars, flat spectrum radio quasars (FSRQ), peak at the lowest energies whereas the faintest, high frequency peaked BL Lac Objects (HBL), show the highest peak frequencies. In between one finds low- and intermediate peaked BL Lac Objects (LBL and IBL respectively).\\
Recent modelling has shown that the SEDs from HBLs to IBLs are reliably reproduced by simple synchrotron self-Compton emission, i.e. the first peak in the spectrum is essentially synchrotron radiation from relativistic primary electrons and the second one arises from Compton upscattering of this synchrotron radiation by the very same electrons \citep{boettcher01, kirk01, tavecchio01, weidinger01}, but especially, when FSRQs are considered the origin of the second hump in the SED is still under debate, whether it is due to comptonized radiation from external broad line region photons, torus photons, or even accretion disk photons entering the jet \citep{dermer01, boettcher03} or due to highly relativistic primary protons within the jet and their synchrotron radiation as well as emerging photo-hadronic cascades \citep{mannheim93, mucke93, boettcher02}. In addition to the stability and formation or the structure of the jet, which cannot be addressed with emission models, this composition of the jet and its difference in FSRQs and HBLs is one of the major
open
questions concerning blazars, and may provide a natural physical explanation of the phenomenological blazar sequence and its recently appearing envelope structure \citep{meyer11}.\\
In this paper we present a time-dependent and fully self-consistent, hybrid emission model for blazars where low energetic electrons and protons are co-accelerated via Fermi-I and Fermi-II processes within a confined region to high energies. Electrons lose energy mainly to the synchrotron and inverse Compton channels, protons radiate synchrotron photons and are subject to photo-meson production, with $\gamma \gamma$-pair-production of highly energetic photons coupling the equations in a non-linear way. Proton collisions of the non-thermal proton distribution are irrelevant at typical densities within the jet and are neglected since no thermal background protons as an additional parameter \citep[e.g.][]{eichmann01} shall be considered.
There have been very recent observations by \citet{agudo01} constraining the $\gamma$-ray emitting site of the jet a few pc away from the black hole. This strongly favours an emission scenario independent of external sources. We are able to model HBLs in the SSC case by setting the number density of injected protons $Q_{0,p^+} \rightarrow 0$ \citep{weidinger01, weidinger02} as well as FSRQs with $Q_{0,p^+} \neq 0$ \citep{weidinger04} with the very same model. Since our model is time-dependent we are able to exploit outbursts of blazars and the timing signatures in different energy bands to narrow down the parameters used in the modelling process and distinguish between leptonic and hadronic dominated jets.\\
In the next section we give a description of our model and its principle properties and assumptions, before we apply it to the recently detected HBL/IBL \mbox{1 ES 1011+496} which, already in the steady state, favours a very high magnetic field being present within the jet making it an outstanding source. In section 4 other, more general, physical implications are discussed.


\section{The model}
\begin{figure}
\begin{center}
\includegraphics[width=\hsize]{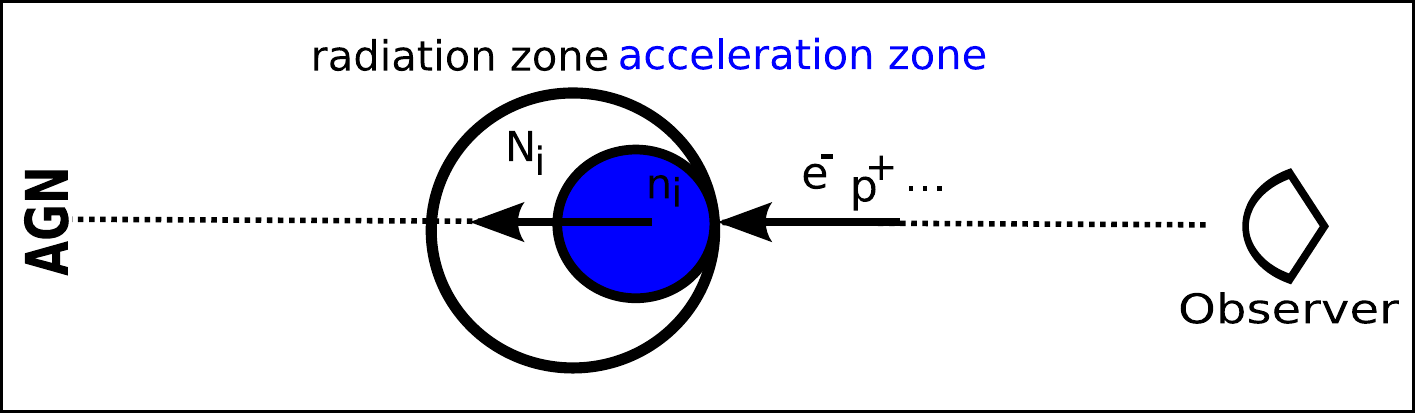}
\caption{Basic model geometry. All escaping particles of the acceleration zone serve as injection for the radiation zone, $i=e^{\pm}$ for electrons/positrons and $i=p^+$ for protons.\label{fig1}}
\end{center}
\end{figure}

\begin{figure}
\begin{center}
\includegraphics[width=\hsize]{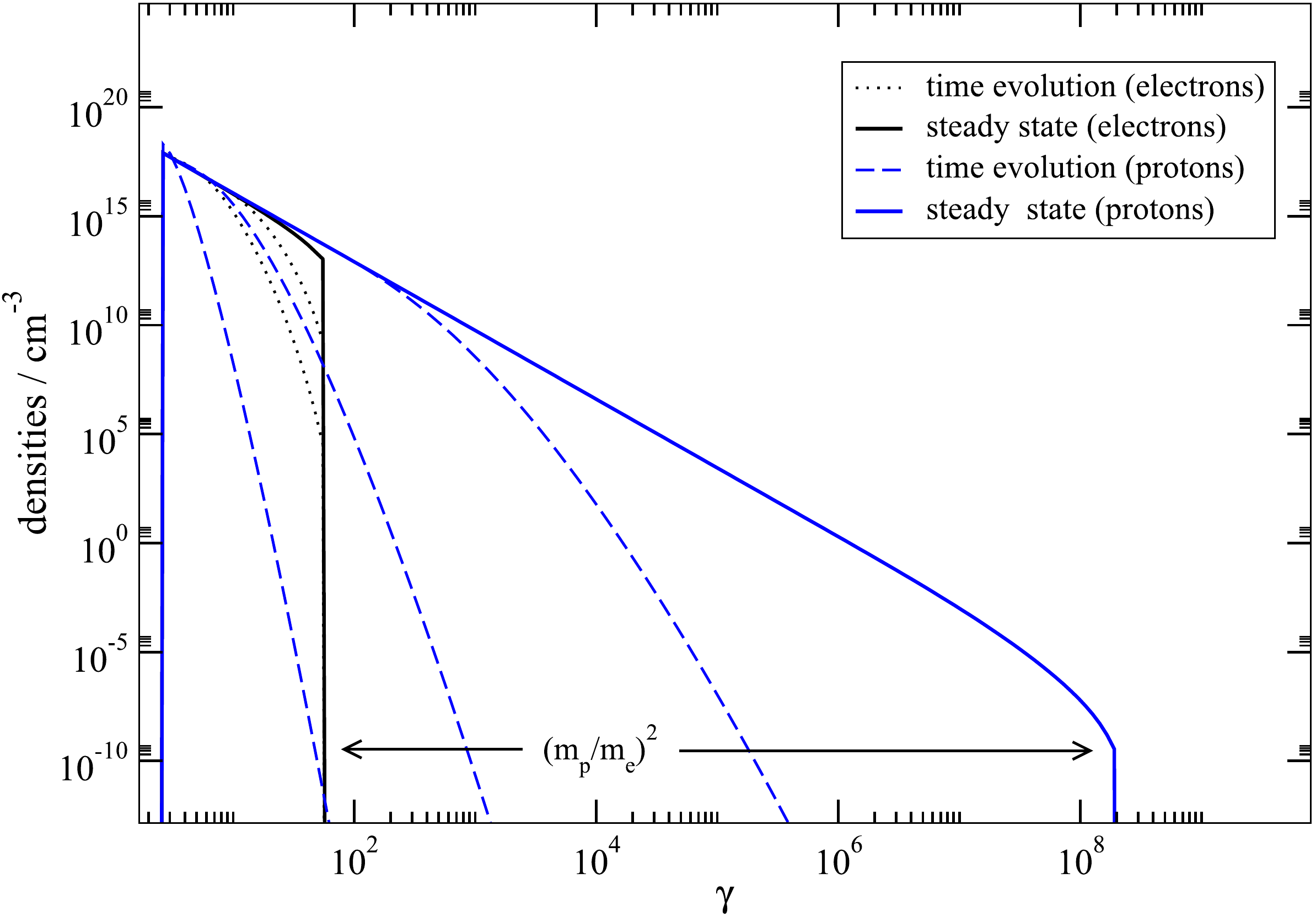}
\caption{Evolution of $p^+$ and $e^-$ of freshly injected particles at $\gamma_{0,i} = 2.5$ in the acceleration zone, assuming $a \rightarrow \infty$. The separation of $\gamma_{max,p^+}\gamma_{max,e^-}^{-1} = m_p^2m_e^{-2}$ arises self-consistently during the simulation. $p^+$ reaches the steady state a factor of $p^+/e^-$ slower than $e^-$, see text for details.\label{fig2}}
\end{center}
\end{figure}

The general modelsetup is assumed to be spherical with an acceleration zone nested inside a radiation zone. As the emitting region (blob) moves down the jet axis towards the observer with a bulk Lorentzfactor $\Gamma$, upstream material (electrons and protons) is picked up and the highly turbulent acceleration zone forms at the edge of the blob. Here both particle species undergo Fermi-I and Fermi-II processes up to relativistic energies with synchrotron losses in a turbulent magnetic field balancing the acceleration with regard to the maximum energy. On the far side of the blob, namely in the considerably larger radiation zone, acceleration is assumed to be inefficient. All calculations are conveniently made in the rest-frame of the blob, the geometry is shown in Figure \ref{fig1}.
Both zones are assumed to be homogeneous and to contain isotropic particle distributions as well as a randomly orientated magnetic field $B$, not to be confused with the helical magnetic field stabilizing the jet against the kinetic pressure of the plasma. The kinetic equations in each zone, one for each particle species $i$, are derived from the relativistic Vlasov equation \citep{schlickeiser02} applying the one-dimensional diffusion approximation in the highly relativistic $p_i = \gamma_i m_i c$ case. Furthermore, the hard-sphere approximation for the spatial diffusion coefficient occurring in the equations is used, see \citet{weidinger01}. We note that when only electrons are picked up into the acceleration zone ($Q_{p^+} \rightarrow 0$) the model reduces to the SSC case described in \citet{weidinger01} and \citet{weidinger02}, which is an extension to \citet{kirk01}.
The acceleration timescales can be translated into the microphysics of the jet, being proportional to the particle's mass in the energy independent case \citep{weidinger04}
\begin{align}
 \label{connection}
 t_{acc,i} = \left(\frac{v_s^2}{4K_{||,i}} +2 \frac{v_A^2}{9K_{||,i}}\right)^{-1} \propto m_i
\end{align}
with the parallel spatial diffusion coefficients $K_{||,i}$, and $v_s$ and $v_A$ as shock and Alfv\'en speeds, respectively, providing the scattering centres mandatory for diffusive shock acceleration (DSA). Hence, Eq. \eqref{connection} can be used to cross check the parameter typical values of the jet's microphysics. Motivated by the typical gyro-timescale, the acceleration timescale for one species is set to be constant and proportional to the particle's mass, while the timescale for the second species results naturally from Eq. \eqref{connection}. Following \citet{vukcevicschlickeiser} this ensures that the particles are accellerated well within the acceleration site. We note that this simple energy independent assumption may slightly underestimate the acceleration efficiency at small values of $\gamma_i$, but guarantees that energy gains at the highest values are not overestimated in terms of DSA.
Eventually all particles escaping the acceleration zone enter the radiation zone downstream the jet. To ensure power-law spectra, as expected from shock acceleration, the escape timescale for the acceleration site is set to be constant and within the order of the acceleration timescale $t_{\text{esc},i} \propto t_{\text{acc},i}$ \citep{kirk01, weidinger01}.
Because of the strong confinement of the particles in the radiation zone, as described in other models, one needs to account for all possible radiation mechanisms, not only the dominating synchrotron losses, as in the acceleration zone.
This requires magnetic fields of $\mathcal{O}(10$ G$)$, typical for hadronic models \citep{mannheim93, boettcher02}, whereas leptonic models typically assume $\mathcal{O}(1$ G$)$ \citep{weidinger01, tavecchio01}. Thus, the same order of equipartition, typically of $\mathcal{O}(10$ \%$)$ in our model, regardless which regime (hybrid or leptonic only) considered, is reached. Hence, self-consistency commands that magnetic fields in blazar models cannot be arbitrarily high without considering non-thermal protons and their radiative output.\\
Because of the model geometry, the acceleration zone is assumed not to contribute to the model SED directly, hence we only solve the kinetic equations for the primary particles. The two relevant kinetic equations for the particles' energy and volume density, which in the isotropic diffusive case take the form
\begin{eqnarray}
 \label{acczone}
\partial_t n_i & = &\partial_{\gamma} \left[( \beta_{s,i} \gamma^2 - t_{{acc,i}}^{-1}\gamma ) \cdot n_i\right]  +\nonumber \\& & + \partial_{\gamma} \left[ [(a+2)t_{{acc,i}}]^{-1}\gamma^2 \partial_{\gamma} n_i\right] + Q_{0,i} - \frac{n_i}{t_{{esc,i}}}
\end{eqnarray}
with $i$ being $i=e^-$ and $i=p^+$ for primary electrons and protons, respectively, and the synchrotron $\beta_{s,i} \propto Bm_i^{-3}$. Consequently $a \propto v_s/v_A$ is the ratio of Shock speed to Alfv\'en velocity, hence denoting the dominance of Fermi-I over Fermi-II processes in our model. Monoenergetic particles are injected into the radiation zone
\begin{eqnarray}
 \label{injection}
 Q_{0,i}(\gamma) & = & Q_{0,i} \delta\left(\gamma-\gamma_{0,i} \right)
\end{eqnarray}
to model the cumulation of particles with densities $Q_{0,i}$ from the upstream direction with Lorentzfactors of $\gamma_{0,i}$ by the blob. The density evolution in the acceleration zone starting with $n_{i} = 0$ at $t_0=0$ can be found in Fig. \ref{fig2}; the parameters are $B=10$ G, $t_{acc}t_{esc}^{-1} = 1.2$, $t_{acc,e} = 3 \cdot 10^{3}$ s and $a \rightarrow \infty$.
As one would expect the slope in Fig. \ref{fig2} for both species is $-(t_{acc}t_{esc}^{-1}+1) = -2.2$, whereas $\gamma_{max,p} = m_p^2m_e^{-2} \gamma_{max,e} = 2\cdot10^8$ since both $\beta_{s,i}$ and $t_{acc,i}$ scale with the particle species' mass and neglecting Fermi-II processes one finds $\gamma_{max,i} = (\beta_{s,i} t_{acc,i})^{-1}$. Note that all these properties result from the self-consistency of our model, which is time-dependent and hence able to reproduce outbursts of blazars by replacing the steady state values of $B$, $t_{acc,i}t_{esc,i}^{-1}$ and/or $Q_{0,i}$ by time-dependent ones.
In the hard sphere approximation $t_{acc,i} \propto m_i \Rightarrow K_{||,i} \propto m_i$ applies, i.e. the particle's mean free path $\langle l \rangle \propto m_i$ as expected. The dependence of $t_{acc}$ on the particle's mass is a rather simple assumption based on the gyro-frequency. A more detailed description would require in-depth knowledge of the turbulent magnetic spectrum and particle scattering properties, which is not available for AGN. Using the hard-sphere approximation as the simpified model assumption to connect the acceleration timescale to the microphysics of the jet, in principle any value of $t_{\text{acc},i}t_{\text{esc},i}^{-1}$ is allowed.
However, we keep in mind that values $t_{\text{acc},i} < t_{\text{esc},i}$ resulting in spectra harder than $s = 2$ are rather hard to motivate in the light of non-relativistic shock acceleration. In this case $s$ is correlated to the compression-ratio $r$ of the shock as $s = (r+2)/(r-1)$ which may be derived from microphysical considerations incorporating bulk plasma speeds and return propabilities of the test-particle. This can be connected to the kinetic equation used (see e.g. \citet{dermeracc}), but is not carried out as the bulk speeds of the underlying plasmoids are not feasible experimentally at this point. Hence, the power-law particle spectra generated in the acceleration zone may be identified as shock acceleration or a closely related process. We note that compression ratios $r > 4$ and, thus, harder particle spectra can be found, for example using non-linear simulations. In addition it should be mentioned that not using the hard-sphere approximation, i.e. $t_\text{acc,i} \rightarrow t_\text{acc,i}(\gamma_i)$, will yield spectra different from power-laws. This case would naturally also require a more sophisticated treatment of $t_\text{esc,i}$.\\
As every particle escaping the acceleration zone enters the radiation zone, the particle spectra of the acceleration zone (see Fig. \ref{fig2}) serve as the injection function for the emission region.\\
Unlike in the acceleration zone, energy gains of the particles are not accounted for at this site. The particles under consideration here will cool because of synchrotron, inverse Compton, and photo-hadronic processes when they radiate. The kinetic equations for electrons, secondary positrons, and protons, as well as the radiative equation are solved self-consistently and time-dependently. Although protons are introduced merely via two new parameters in comparison to the self-Compton case described in \citet{weidinger01}, namely the injected number density $Q_{0,p^+}$ and the corresponding $\gamma_{0,p^+}$, this has a major effect in the radiation zone, especially when time-dependency is considered, since photo-hadronic processes emerge as a significant contribution. This contribution has an impact on the radiative equation and, more importantly couples the kinetic equation of the electrons/protons to the radiative equation in a non-linear way as free pair-processes will play a major role as well, in the
case of non vanishing proton densities.\\
For the protons the kinetic equation in the radiation zone hence yields
\begin{eqnarray}
 \label{radpro}
 \partial_t N_{p^+} & = & \partial_{\gamma} \left[\left(\beta_{s,p^+}\gamma^2 + P_{\text{p}\gamma}(\gamma)\right) N_{p^+}\right] + b \frac{n_{p^+}}{t_{esc,p^+}} - \frac{N_{p^+}}{t_{esc,rad,p^+}}
\end{eqnarray}
with the parameter $b < 1$ ensuring particle conservation. Following the discussion of \citet{blumenthal01} and \citet{berezinsky01} the losses due to photo-meson production $P_{\text{p}\gamma}$ can be neglected against the dominating synchrotron losses for most blazars, substantially lowering computational costs. Highly relativistic protons with $\gamma_{p^+} > 10^5$ are subject to photo-meson production with the radiation field within the emitting region, mainly pion production
\begin{equation}
 \label{photopion}
 p^+ + \gamma \rightarrow n_0 ~\pi^0 + n_+ ~\pi^+ + n_- ~\pi^- + \mbox{neutrons}.
\end{equation}
The produced pions are unstable particles and decay into stable $e^{\pm}$ (and $\gamma$s) via the muon channel, thereby producing neutrinos of flavours
\begin{eqnarray}
 \label{chains}
 \pi^{+} & \rightarrow & \mu^{+} + \nu_{\mu} \rightarrow e^{+} + \nu_e + \nu_{\mu}, \nonumber \\ \pi^{-} & \rightarrow & \mu^{-} + \bar{\nu}_{\mu} \rightarrow e^{-} + \bar{\nu}_e + \bar{\nu}_{\mu}, \nonumber \\ \pi^0 & \rightarrow & \gamma + \gamma.
\end{eqnarray}
To calculate the production rate of stable particles from pion production we use the \citet{kelner01} parametrization (including their erratum) of the SOPHIA Monte Carlo results \citep{mucke00} for these processes. Hence, we assume the photo-hadronic interactions to be instantaneous compared to the synchrotron loss timescale, i.e. we do not account for synchrotron losses and radiation of the intermediate particles. In high magnetic fields this error remains small \citep{mucke93}. When considering fairly soft power-law particle spectra, as in typical blazars, Bethe-Heitler pair-production as considered in \citet{mastichiadis} can be neglected since its contribution is orders of magnitude below the $p^+$-synchrotron radiation, see e.g. \citet{boettcher06}. Even though the Bethe-Heitler process sets in at lower energies, the resulting pairs in magnetic fields as low as considered here will not produce a significant contribution to the SED,  although this might be different for other blazar flavours e.g \citet{murase}. Nevertheless, photon quenching and runaway secondary production \citep{mastichiadis2} will occur in our model as well through the photo-meson channel, given the corresponding parameters (i.e. $p^+$-densities).\\
The resulting $\gamma$s from the $\pi^0$ decay as well as synchrotron radiation of the secondary $e^{\pm}$ from the $\pi^{\pm}$ decay will partially be in the optically thick regime. Pair production
\begin{equation}
 \label{ggpp}
 \gamma + \gamma \rightarrow e^+ + e^-
\end{equation}
will occur, mainly with the synchrotron radiation field of the primary accelerated electrons. Therefore, electromagnetic cascades will emerge until the resulting synchrotron radiation is visible in the optically thin regime for pair-production. These cascades have a major non-linear effect on the kinetic equations of the electrons, protons and photons.\\
The kinetic equations for the electrons and positrons in the radiation zone thus are
\begin{eqnarray}
 \partial_t N_{e^-} &= &\partial_{\gamma}\left[\left(\beta_{s,e} \gamma^2 + P_{{IC}}(\gamma)\right) \cdot N_{e^-} \right] - \frac{N_{e^-}}{t_{{rad,esc,e}}} \nonumber \\&&+ Q_{{pp}}(\gamma) + Q_{{p}\gamma^-}(\gamma) + b\frac{n_{e^-}}{t_{{esc,e}}} \label{radele}\\
 \partial_t N_{e^+} &= &\partial_{\gamma}\left[\left(\beta_{s,e} \gamma^2 + P_{{IC}}(\gamma)\right) \cdot N_{e^+} \right] - \frac{N_{e^+}}{t_{{rad,esc,e}}} \nonumber \\&& + Q_{{pp}}(\gamma) + Q_{{p}\gamma^+}(\gamma)~.\label{radpos}
\end{eqnarray}
No accelerated primary positrons are assumed ($n_{e^+} = 0$). The inverse Compton loss rate ($P_{IC}$) for isotropically distributed $e^{\pm}$ is calculated exploiting the full Klein-Nishina cross section in Eq. (\ref{iclosses}) from e.g. \citet{blumenthal70},
\begin{eqnarray}
 \frac{\mathrm{d}N(\gamma,\alpha_1)}{\mathrm{d}t~ \mathrm{d}\alpha} =& \frac{2 \pi r_0^2 c}{\alpha_1 \gamma^2} &\bigg[ 2 q\ln q+(1+2q)(1-q)\nonumber \\&&+\frac{1}{2} \frac{(4 \alpha_1 \gamma q)^2}{(1+4\alpha_1 \gamma q)}(1-q)  \bigg]~, \label{fullkn}
\end{eqnarray}
with $r_0 = e^2/(mc^2)$, $h\nu = \alpha mc^2$, and $q={\alpha}/(4\alpha_1 \gamma^2(1-\alpha/\gamma))$ as the scattering parameter, i.e. $\alpha_1$ being the incident photon's energy, the scattering geometry allows for $\alpha_1<\alpha\leq 4\alpha_1\gamma^2/(1+4\alpha_1\gamma)$ which is used to determine $\alpha_{max}$
\begin{eqnarray}
 \label{iclosses}
 P_{{IC}}(\gamma) &= m^3c^7h&\hspace{-0.2cm} \int_{0}^{\alpha_{max}}\mathrm{d}\alpha \alpha \int_0^{\infty}\mathrm{d}\alpha_1 \left(N_{{ph}}(\alpha_1) \frac{\mathrm{d}N(\gamma,\alpha_1)}{\mathrm{d}t~\mathrm{d}\alpha}\right)~.
\end{eqnarray}
The model discussed here does not assume external photon fields, the consideration of anisotropic photon fields as in \citet{dermer09} and \citet{hutter11} is, therefore, not necessary.

The pair-production rate in Eqs. (\ref{radele}) and (\ref{radpos}) for photons of energy $\alpha_1$ and $\alpha_2$ respectively is calculated using the approximation
\begin{eqnarray}
 Q_{pp}(\gamma) & = & \frac{3}{32}c\sigma_T  \int_{\gamma}^{\infty}{\mathrm{d}\alpha_1} \int_{\alpha_{min}}^{\infty}{\mathrm{d}\alpha_2}\nonumber \\&& \left[ \frac{4\alpha_1^2}{\gamma(\alpha_1-\gamma)}\ln{\left(\frac{4\alpha_2\gamma(\alpha_1-\gamma)}{\alpha_1}\right)}-8\alpha_1\alpha_2\right.\nonumber\\ && \left.+\frac{2(2\alpha_1\alpha_2-1)\alpha_1^2}{\gamma(\alpha_1-\gamma)}\right.\left.-\left(1-\frac{1}{\alpha_1\alpha_2}\frac{\alpha_1^4}{\gamma^2(\alpha_1-\gamma)^2} \right) \right] \label{qpair}
\end{eqnarray}
of \citet{boettcher04} for isotropically distributed photons for the blob's photon field $N_{ph}$ with itself, with $\alpha_{min}^{-1} = 4\gamma(\alpha_1-\gamma)\alpha_1^{-1}$. Which has an error in energy of less than three percent over the whole electron distribution compared with the absorbed photons Eq. (\ref{app}) and is numerically much more stable than the full production rate (Eq. (26) in their paper). The injection rate of secondary $e^{\pm}$ hence is
\begin{eqnarray}
 Q_{p\gamma^{\pm}}(\gamma) & = &  \frac{m_e}{m_p} \int_0^{\infty}{\mathrm{d}\nu}\int_1^{\infty}{\mathrm{d}\gamma^{'}}\frac{N_{p^+}(\gamma^{'})}{\gamma^{'}}N_{ph}(\nu)\cdot\nonumber\\&& \cdot \Phi_{\pm}\left(\kappa,\frac{m_e\gamma}{m_p\gamma^{'}}\right)~,~~\kappa = \frac{4h\nu\gamma^{'}}{m_e m_pc^4}\label{pgamma}
\end{eqnarray}
with the corresponding $\Phi_{\pm}$ parameterization function of \citet{kelner01}.\\
The photon distribution in the radiation zone, which is eventually beamed towards the observer, reads
\begin{eqnarray}
 \partial_t N_{ph} & = & R_s(\nu) + R_c(\nu) + R_{\pi^0}(\nu) \nonumber \\ && - c \left(\alpha_{{SSA}}(\nu) + \alpha_{{pp}}(\nu) \right) N_{ph} - \frac{N_{ph}}{t_{{ph,esc}}}\label{radpho}
\end{eqnarray}
taking all mentioned processes into account. The photon escape timescale is the light crossing time. With the synchrotron emissivity $R_s$ as the Melrose approximation
\begin{eqnarray}
 R_s(\nu) & = & 1.8 \frac{\sqrt{3}e^3 B_{\bot}}{h \nu m c^2} \int{\mathrm{d}\gamma N_e(\gamma, t) \left( \frac{\nu}{\nu_c(\gamma)}\right)^{\frac{1}{3}} e^{-\frac{\nu}{\nu_c(\gamma)}}}~,\nonumber\\&&\nu_c(\gamma) = \frac{3\gamma^2 e B_{\bot}}{4\pi m c}~, \label{rs}
\end{eqnarray}
see \citet{weidinger01} and the inverse Compton production rate
\begin{eqnarray}
R_{c}(\nu) & = \int \mathrm{d}\gamma\, N_{e}(\gamma) \cdot \int \mathrm{d}\alpha_1 &
\left[N_{{ph}}(\alpha_1)\frac{\mathrm{d}N(\gamma,\alpha_1)}{\mathrm{d}t~\mathrm{d}\alpha}\right. \nonumber \\ && \left.- N_{{ph}}(\alpha)
\frac{\mathrm{d}N(\gamma,\alpha)}{\mathrm{d}t~\mathrm{d}\alpha_1}\right]\label{rc}
\end{eqnarray}
using the full Klein-Nishina cross section Eq. (\ref{fullkn}). The photons from the $\pi^0$ decay are calculated analogously to Eq. (\ref{pgamma}), again with the corresponding $\Phi_0$ parametrization of the SOPHIA results:
\begin{eqnarray}
 R_{\pi^0}(\nu) & = & \frac{h}{m_pc^2} \int_0^{\infty}{\mathrm{d}\nu^{'}}\int_1^{\infty}{\mathrm{d}\gamma}\frac{N_{p^+}(\gamma^{'})}{\gamma}N_{ph}(\nu^{'})\cdot \nonumber \\ && \cdot \Phi_{0}\left(\kappa,\frac{h\nu}{\gamma m_pc^2}\right)~.\label{rp0}
\end{eqnarray}
In optically thick regimes the photon field is absorbed either in the low energies as a result of synchrotron self absorption by the emitting electrons / positrons themselves ($\alpha_{SSA}$) for which the monochromatic approximation \citep{weidinger02} is used, or in the VHEs because of $e^\pm$-pair-production. The photon annihilation coefficient for $e^{\pm}$-production is calculated using the exact result of \citet{coppi01} for isotropic pair plasmas,
\begin{eqnarray}
 \label{app}
 \alpha_{PP}(\nu) & = & \int_0^{\infty}{\mathrm{d}\nu'}N_{Ph}(\nu')\int_{-1}^{\mu_{max}}{\mathrm{d}\mu}\frac{1-\mu}{2}\sigma(x,\mu)
\end{eqnarray}
with $\mu_{max}=\max{(-1,1-2x^{-1})}$, where $x=\alpha\alpha_1$ and
\begin{eqnarray}
 \label{ppcs}
 \sigma(x,\mu) & = & \frac{3\sigma_T}{8x(1-\mu)}\left[\left(3-\left(1-\frac{2}{x(1-\mu)}\right)^2\right)\cdot \right.\nonumber\\&& \cdot \ln{\left(1+\frac{\sqrt{1-\frac{2}{x(1-\mu)}}}{1-\sqrt{1-\frac{2}{x(1-\mu)}}}\right)}\nonumber\\ & & \left.-2\sqrt{1-\frac{2}{x(1-\mu)}}\left(1+\frac{2}{x(1-\mu)} \right)\right]
\end{eqnarray}
being the full cross section for $\gamma\gamma$-pair-annihilation. The photon distribution $N_{ph}$ is transformed into the observer's frame using the beaming pattern for isotropic photon distributions in a sphere while accounting for the redshift in order to achieve the model SED of the considered blazar.\\\\
Within the model assumptions it is possible to explain the VHE peak in the typical blazar spectra either as inverse Compton upscattering of synchrotron photons or as proton synchrotron radiation consistently accompanied by cascaded radiation from the photo-hadronic interactions, just depending on the chosen parameters. We note that the $p^+$-synchrotron radiation important for the occurrence of the second peak in a blazar's SED when considering the hybrid-case naturally becomes dominant at proton densities considered for substantial photo-meson production. Thanks to the numerical treatment of the processes not relying on Monte Carlos processes, we can compute inter-band lightcurves even in the hadronic case making the potential information from outbursts of blazars accessible for physical interpretations.
In the hadronic case the inter-band lightcurves exhibit more complex features (highly dependent on the chosen set of parameters) like orphan flares, typical time-lags, etc., unknown to purely leptonic models.

\section{Application to 1 ES 1011+496}
Despite prominent absorption lines used to pin down the redshift to $z = 0.212$ the blazar 1 ES 1011+496 was originally classified as a high frequency peaked BL Lac object \citep{albert01}. The VHE emission of 1 ES 1011+496 was discovered with the MAGIC Air-Cherenkov telescope in 2007, and it was the most distant blazar observed in $\gamma$ rays while down to the present day this is 3C 279.
However, the true nature of this blazar seems to be quite unclear compared to well studied objects like Mkn 501, PKS 1218+304 or 3C 279 \citep{veritas, weidinger02, boettcher02}. The first and only multiwavelength observation, including MAGIC in the VHEs, Swift in X-rays and KVA in the optical band was in 2008 \citep{reinthal01}.
During this campaign the emission of 1 ES 1011+496 was more or less steady with slight spectral variability in the optical and X-rays having no counterpart in the $\gamma$ rays \citep{reinthal01}, which of course may be due to the low observed flux and sensitivity of the instrument. The long-term optical monitoring of 1 ES 1011+496 also indicates variability consistent with synchrotron emission and magnetic field fluctuations \citep{boettcher05}.
The multiwavelength data of \mbox{1 ES 1011+496} will now be used to study the properties of this interesting blazar in more detail with our model. The one-year Fermi-LAT butterfly is considered as an upper limit, since the MWL data represents a relatively low flux level of 1 ES 1011+496.

\subsection{The spectral energy distribution} \label{bozomath}
\begin{figure}
\begin{center}
\includegraphics[width=\hsize]{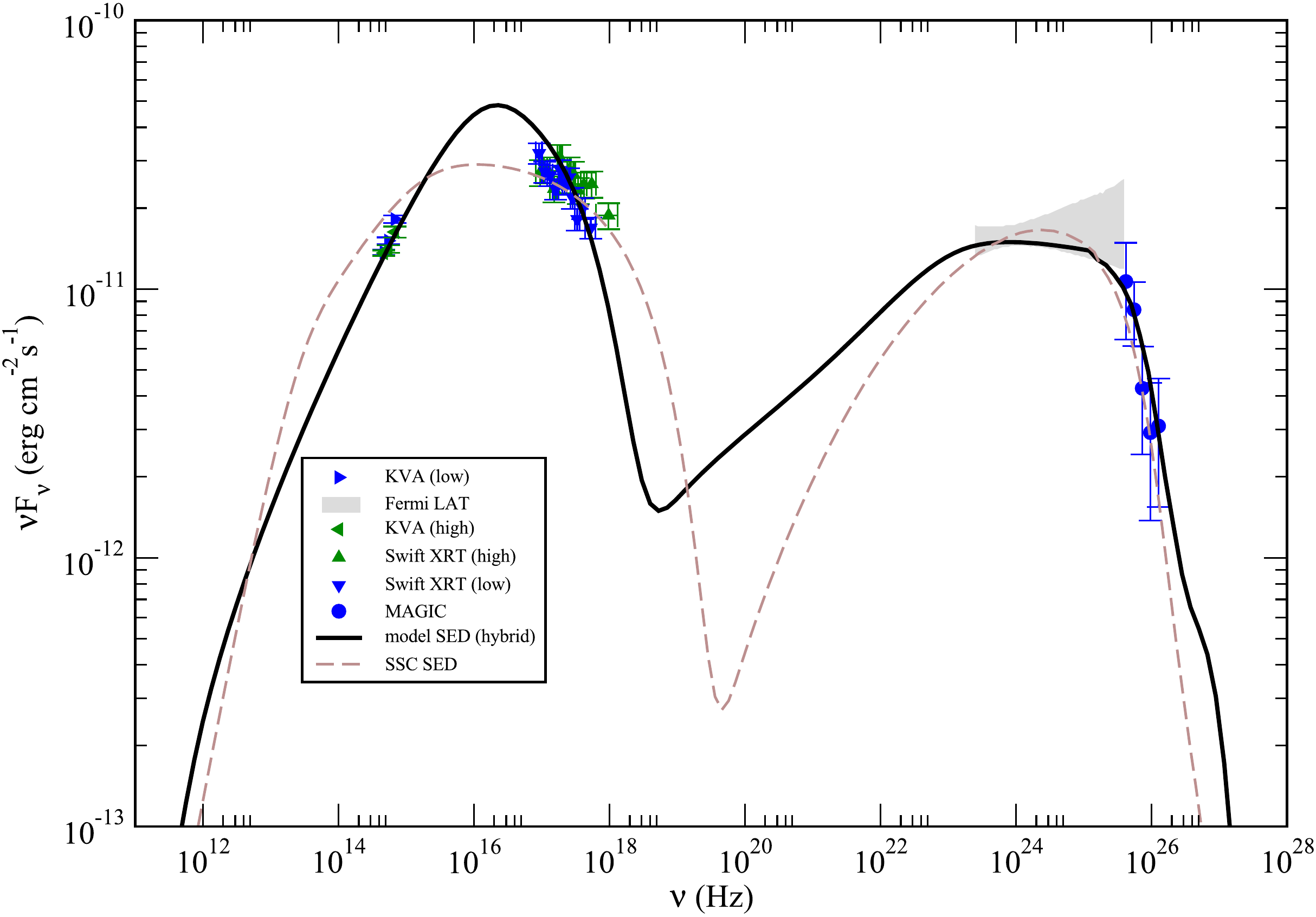}
\caption{Simultaneous data from \citet{reinthal01} with KVA in the optical, Swift XRT in the X-rays, and MAGIC in the VHEs as measured. Blue symbols indicate the low emission state used in the modelling. The high state inferred from the slight variability is shown in green. The grey butterfly represents the first-year catalogue spectrum of Fermi LAT \citep{fermifirst}. The grey (online: brown) dashed curve displays the modelling attempt with a low magnetic field and no protons present in the jet (Table 1a); the black solid curve is due to the modelling with high $B$-field co-accelerated protons (Table 1b). The model SEDs have been EBL-absorbed using the model of \citet{primack05}\label{fig3}}
\end{center}
\end{figure}

In Figure \ref{fig3} the measured multiwavelength spectrum from \citet{reinthal01} and two model SEDs are shown. We note that we use the EBL absorbed data points in this paper. The parameters of the models can be found in Table \ref{tbl1}a and \ref{tbl1}b.
The rather hard spectrum in the optical combined with a photon index of $s = 2.32$ in the Swift XRT band makes it difficult to explain the SED of 1 ES 1011+496 in terms of a self-consistent SSC-Ansatz already in the steady state (brown dashed line in Fig. \ref{fig3}). The best parameters in the self-Compton limit, i.e. a low magnetic field unable to confine non-thermal protons within the considered region, are shown in Table \ref{tbl1}a. A spectral index of $2.2$ for the electrons with a magnetic field of $0.18$ G can neither reproduce the narrow synchrotron peak nor the spectral properties when computed self-consistently.
Of course one can understand the broadband SED of 1 ES 1011+496 with a purely leptonic model, see e.g. \citet{reinthal01} itself, but in those cases the underlying electron spectra do not arise in the modelling process but are adhoc assumptions which spectral indices and breaks requiring physical motivation put in by hand. Consistent cooling breaks fail to reproduce the synchrotron peak at low $B$-fields, automatically yielding a relatively high magnetic field being present within the jet.
In such a case $r_{gyr} \ll R_{rad}$ is fulfilled even for relativistic protons making them effective emitters within the jet, hence being co-accelerated in the acceleration zone, leading to a hybrid spectrum of this particular blazar, see Fig. \ref{fig3}. The parameters used can be found in Table \ref{tbl1}b.
\begin{table*}
\caption{Parameters found in the modelling process to the multiwavelength data of 1 ES 1011+496 a) using low magnetic fields, b) high magnetic fields thus confined highly relativistic protons.}
\label{tbl1}
\centering
\begin{tabular}{l c c c c c c c c c c }
\hline\hline
Model & $Q_{0,p^+}$ (cm$^{-3}$)& $\gamma_{0,p^+}$ & $Q_{0,e^-}$ (cm$^{-3}$) & $\gamma_{0,e^-}$ &
$B$ (G) & $t_{acc,e}$ (s) & $R_{blob}$ (cm) & $t_{acc}/t_{esc}$ & $a$ & $\delta$\\
\hline
a) & $0$ & $-$ & $7.50\cdot 10^4$ & $868$ & $0.18$ & $3.5 \cdot 10^{4}$ & $8.00 \cdot 10^{15}$ & $1.2$ & $1.0 \cdot 10^3$ & $44$ \\
b) & $1.55 \cdot 10^8$ & $600$ & $3.78\cdot 10^7$ & $3400$ & $8.0$ & $3.7 \cdot 10^{2}$ & $1.75 \cdot 10^{15}$ & $1.3$ & $20$ & $36$\\
\hline
\end{tabular}
\end{table*}
In a magnetic field of $8.0$ G the power-law electrons (with the corresponding $t_{acc}$) cannot exceed Lorentzfactors of $\gamma_e \approx 2.5 \cdot 10^4$ because of synchrotron losses, which together with the relatively high $\gamma_{0,e^-}$ leads to a narrow peaked $\nu F_{\nu}$-spectrum in the X-ray band, explaining the SED of 1 ES 1011+496 there. The Fermi-LAT data on other blazars also suggests that a high $\gamma_{0e^-}$ is present, e.g. \citet{reinthal01, weidinger01}. These values of $\gamma_{0i}$ are expected since Fermi acceleration requires pre-accelerated, non-thermal particles to be efficient. Hence the ratio of $\gamma_{0p^+}/\gamma_{0e^-}$ depends on this process and might give hints on the nature of this mechanism, e.g. trapping of low energetic $e^-$. In magnetic fields as high as inferred here, inverse Compton scattering of synchrotron photons is negligible. The second peak in the spectrum thus consists of synchrotron radiation of highly relativistic protons, consistently accelerated to $\gamma$ $\approx 8.7 \cdot 10^{10}$ in the model,
 as well as pair cascades arising from photo-hadronic interactions of these protons with the synchrotron photons.
With the parameters shown in Table \ref{tbl1}b the measured MWL spectrum of 1 ES 1011+469 is explained well, introducing no new parameters other than the number of injected low energy protons and their injection energy (or the electron/proton ratio and - energy ratio) as compared to the (simple) SSC-limit. The injected luminosity (i.e. the jet luminosity as the blob moves along) is $L_p = 3.0 \cdot 10^{42}$ erg s$^{-1}$ in hadrons, assuming a vanishing angle to the line of sight. This is significantly below the Eddington limit for the black hole assumed to be present in 1 ES 1011+496 \citep{fan07}. Proton spectra extending down to $\gamma_{0p^+} = 1$ would dramatically increase the kinetic luminosity without significant radiative output. The photon indices of the second peak are not mapped to the synchrotron spectrum of the radiating protons directly, since it also contains redistributed radiation (see Fig. \ref{fig4}) hardening the VHE peak. The spectral indices of the underlying particle species ($s = 2.25$ for
both) is consistent with diffusive shock acceleration at a strong shock.

\subsection{Intrinsic spectrum of 1 ES 1011+496}
\begin{figure}
\begin{center}
\includegraphics[width=\hsize]{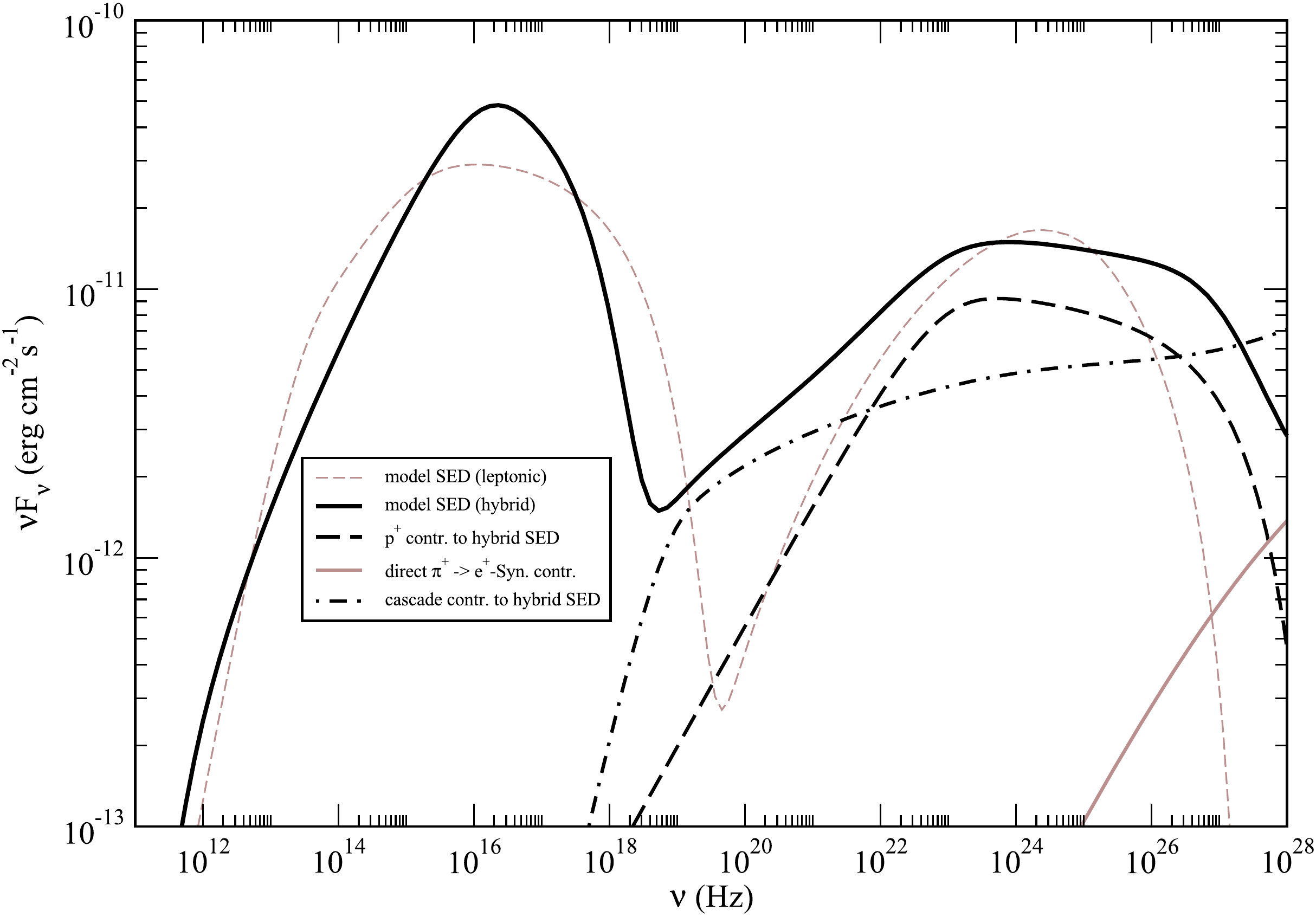}
\caption{Intrinsic SEDs (unaffected by EBL absorption) of 1 ES 1011+496 (Fig. \ref{fig3}) as inferred by the model, parameters in Table \ref{tbl1}. The solid black line shows the hybrid SED with the individual components: proton synchrotron emission (dashed black line) and redistributed synchrotron emission by $e^{\pm}$-pairs initiated by photo-hadronic interactions (dash-dotted black line) only observable through pair-cascades. The direct contribution of stable end-products from photo-pion production are negligible in the observable regime peaking above $10^{27}$ Hz opaque to $\gamma \gamma$-pair-production, see solid brown line ($\pi^+ \rightarrow e^+$; other processes are below $10^{-13}$ erg cm$^{-2}$ s$^{-1}$) and only affect the SED indirectly via $\gamma\gamma$-pair-production of the $n$th generation.\label{fig4}}
\end{center}
\end{figure}

To resolve the second peak in the spectrum of 1 ES 1011+496 it is much more convenient to take a look at the intrinsic SED unaffected by the EBL, shown in Fig. \ref{fig4}. As already mentioned the first peak in the SED consists of synchrotron radiation of the primary electrons within the radiation zone. The situation for the second hump, however, is slightly more complex.
The relevant contributions to the second peak of 1 ES 1011+496 are proton synchrotron photons of the highly relativistic primary $p^+$ with Lorentzfactors up to $\gamma_p \approx 10^{10}$ and cascade radiation. The latter is synchrotron radiation of the stable products arising from photo-meson production which is redistributed from the optically thick regime for pair-production until observable. As one can infer from Fig. \ref{fig4} (solid brown line) the direct contribution of $p\gamma$ interactions is negligible thanks to the dominance of the proton synchrotron peak. That is also why Bethe-Heitler pair-production with a lower threshold than photo-hadronic processes can be neglected against the proton synchrotron emission in magnetic fields of $\mathcal{O}(10$ G$)$ required to confine the protons within the emitting region of a typical blazar. From Fig. \ref{fig4} it is also clear that the maxima of the first generation radiation of $e^{\pm}$ and $\gamma$s from $p\gamma$-interactions are above $10^{28}$ Hz.

It should be noted that the attenuation of the spectrum by the EBL may have secondary effects. It has been discussed in the literature \citep[cf.][]{neronov10} that EBL absorption yields electron-positron pairs, which can upscatter CMB photons to GeV energies. The detection or non-detection of these GeV photons by Fermi, can give rise to the determination of the the intergalactic magnetic field (IGMF; assuming that diffusion of electron-positron pairs in the IGMF is the only process preventing the Compton scattering). The integrated flux above $1$ TeV in the hybrid scenario is an order of magnitude higher than the hard spectra assumed in \citet{neronov10} and as the redshift of \mbox{1 ES 1011+496} is comparable, we find a crude lower bound for the IGMF of $\mathcal{O}(10^{-15}$ G$)$, which has been reported in other analyses, e.g. \citet{tavecchioIGMF}. This is an extremely rough estimate, since different regions of the EBL are touched in the hybrid case compared to purely leptonic spectra which cut off at lower energies, additionally leading to deviating IC cooling timescales. It should not be neglected here that plasma effects (namely the pair-beam instability) with substentially smaller mean free paths are passionatly discussed as the dominating cooling process for TeV beams, not producing a radiative signature in the Fermi LAT band, even in the absence of an IGMF \citep{schlickeiser13, broderick12}. Furthermore the pair creation probability inside the jet, but outside the actual source region has not been modelled, which may substantially alter the TeV spectrum of a blazar injected into the intergalactic medium.
\subsection{Variability}
\begin{figure}
\begin{center}
\includegraphics[width=\hsize]{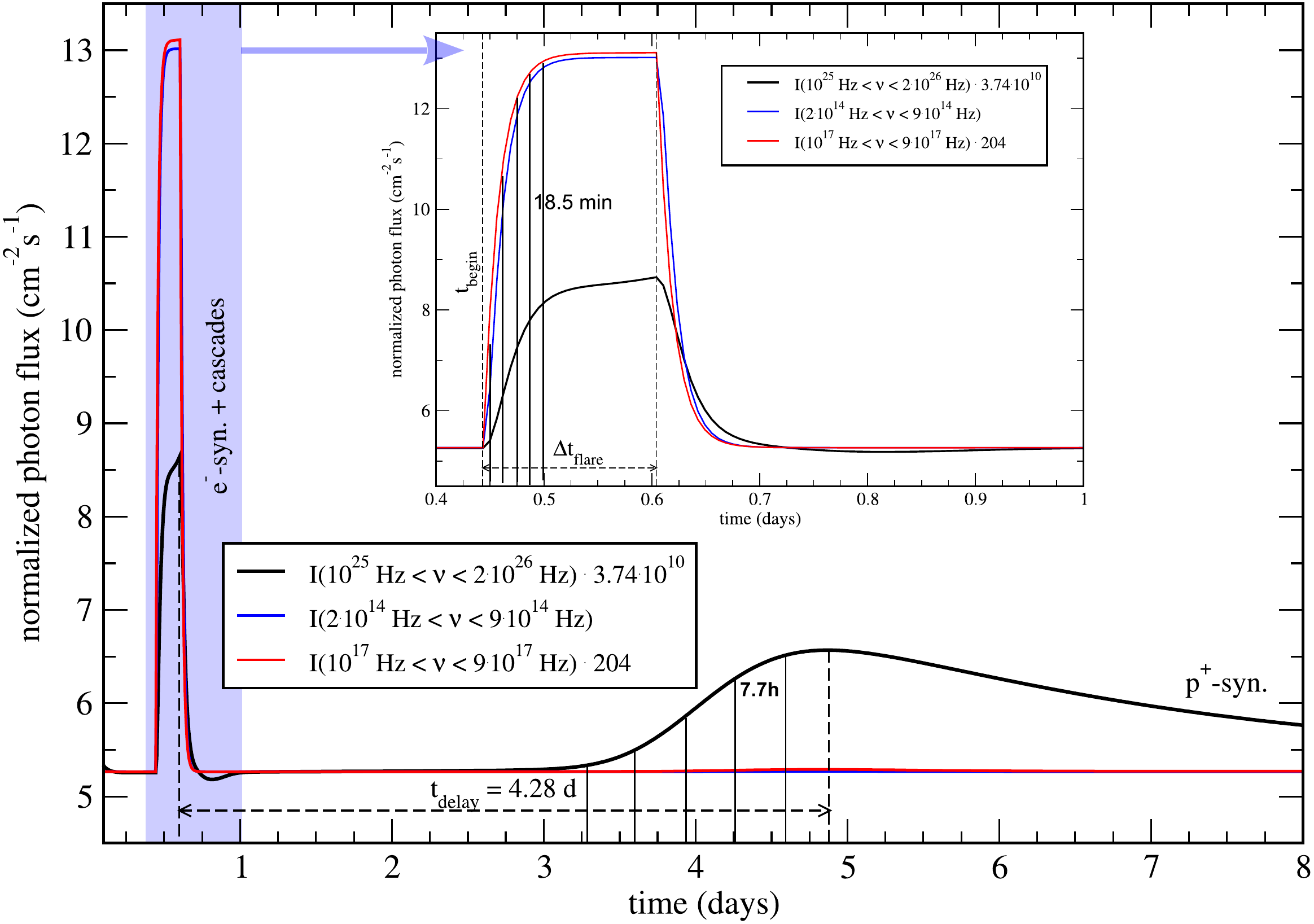}
\caption{Multiband lightcurve during the assumed scenario of density fluctuations along the jet axis explained in the text. The inset shows the definition of the snapshots during the first outburst in Fig. \ref{fig6}. The timescale of this flare corresponds to the synchrotron timescale of the electrons with a slight delay in the gamma rays caused by the cascades that need to be built up first. With a delay of about $4.28$ days a second, orphan flare occurs in the VHEs when the freshly injected protons have been accelerated to high Lorentzfactors to radiate efficiently. The timescale here corresponds to the synchrotron loss timescale of protons.\label{fig5}}
\end{center}
\end{figure}

\begin{figure}
\begin{center}
\includegraphics[width=\hsize]{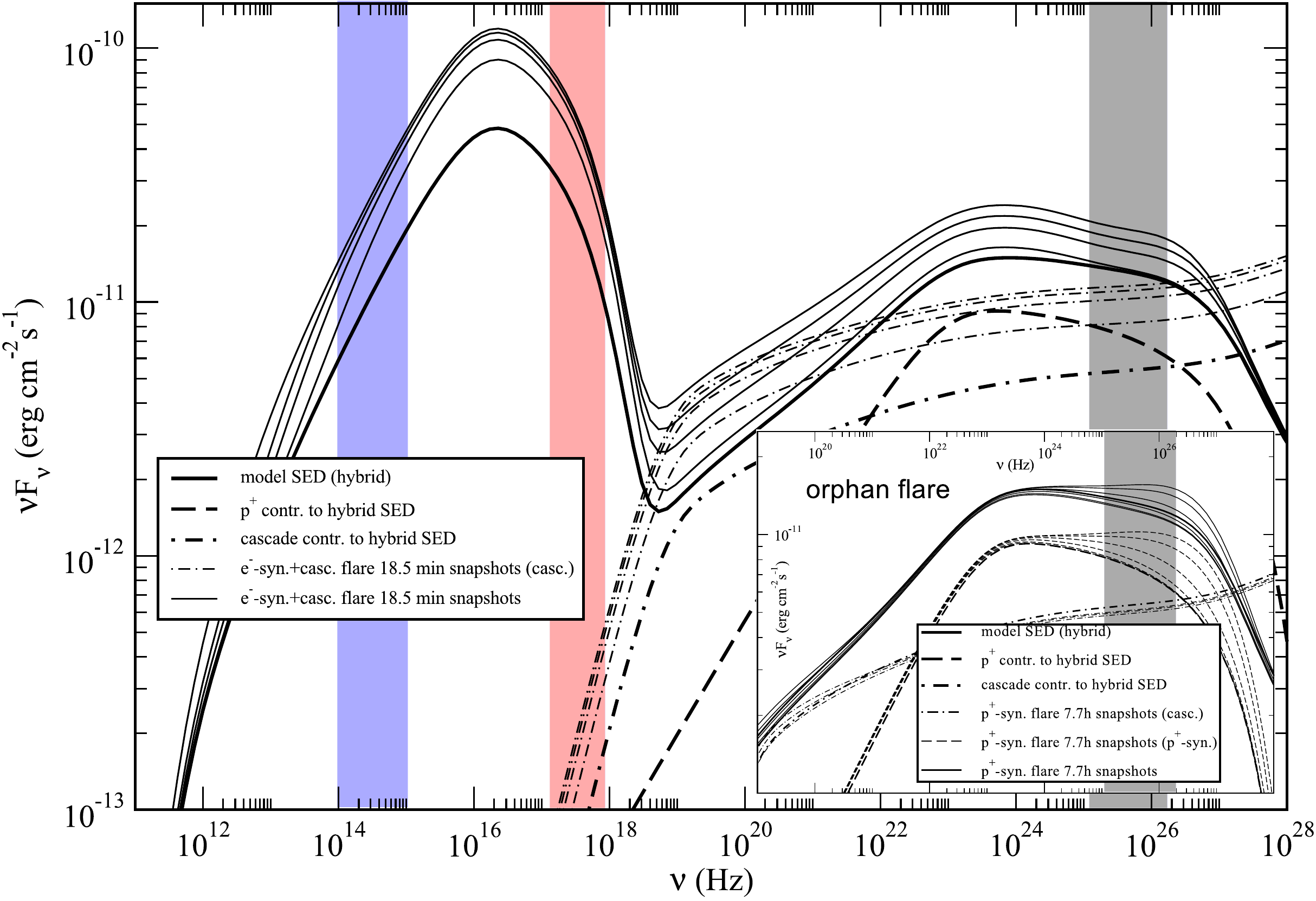}
\caption{Spectral evolution ($18.5$ min snapshots) during the imposed flaring scenario of 1 ES 1011+496 (intrinsic spectrum). The VHE peak rises with the synchrotron peak because of the enhanced seed photons for hadronic interactions and the redistribution via $\gamma\gamma \rightarrow e^{\pm}$. The inset shows the second, orphan flare in the gamma rays (see also Fig. \ref{fig5}) due to enhanced proton synchrotron radiation occurring with a delay, since the protons need time to accelerate to high energies.\label{fig6}}
\end{center}
\end{figure}

Yet another advantage of the numerical approach is that it is not only possible to use exact cross sections including all non-linear interactions, but that a time-dependent treatment is feasible and variability becomes accessible even in the lepto-hadronic case.
The structure of the VHE peak in the hybrid scenario including leptonic and hadronic emission will involve various timescales of variability which can be investigated with our model. In this section we introduce a possible flaring scenario for 1 ES 1011+496, where more primary $e^-$ and $p^+$ are injected into the steady state acceleration zone for a certain amount of time $\Delta t$, i.e.
\begin{eqnarray}
  Q_{0,i}(t) = Q_{0,i} \cdot \left\{\begin{array}{*{2}{l}}
 x_i & , ~ t_{begin} < t <  t_{begin} + \Delta t \\
 1 & , ~ \mbox{otherwise.} \label{flare}\\
\end{array}
\right.
\end{eqnarray}
This scenario might occur within the jet of a blazar as the blob moves through strong density fluctuations along the axis of the outflow. Unlike the SSC models, we expect long time-lags and different timescales to occur in the multiband lightcurve serving as a hadronic fingerprint of the considered blazar. All timescales and times are given in the observer's frame.
\\Figure \ref{fig5} shows the response in the optical (blue), X-rays (red), and gamma rays (black) when (arbitrarily) setting $\Delta t = 3.86$ h and $x_e = 2.5$, $x_p = 8.5$. Of course these parameters need to be interpreted and will be different observing an actual outburst of 1 ES 1011+496 in various energy bands. The definition of the energy ranges in the multiband lightcurves of Fig. \ref{fig5} can be found in Fig. \ref{fig6}.
\begin{figure}
\begin{center}
\includegraphics[width=\hsize]{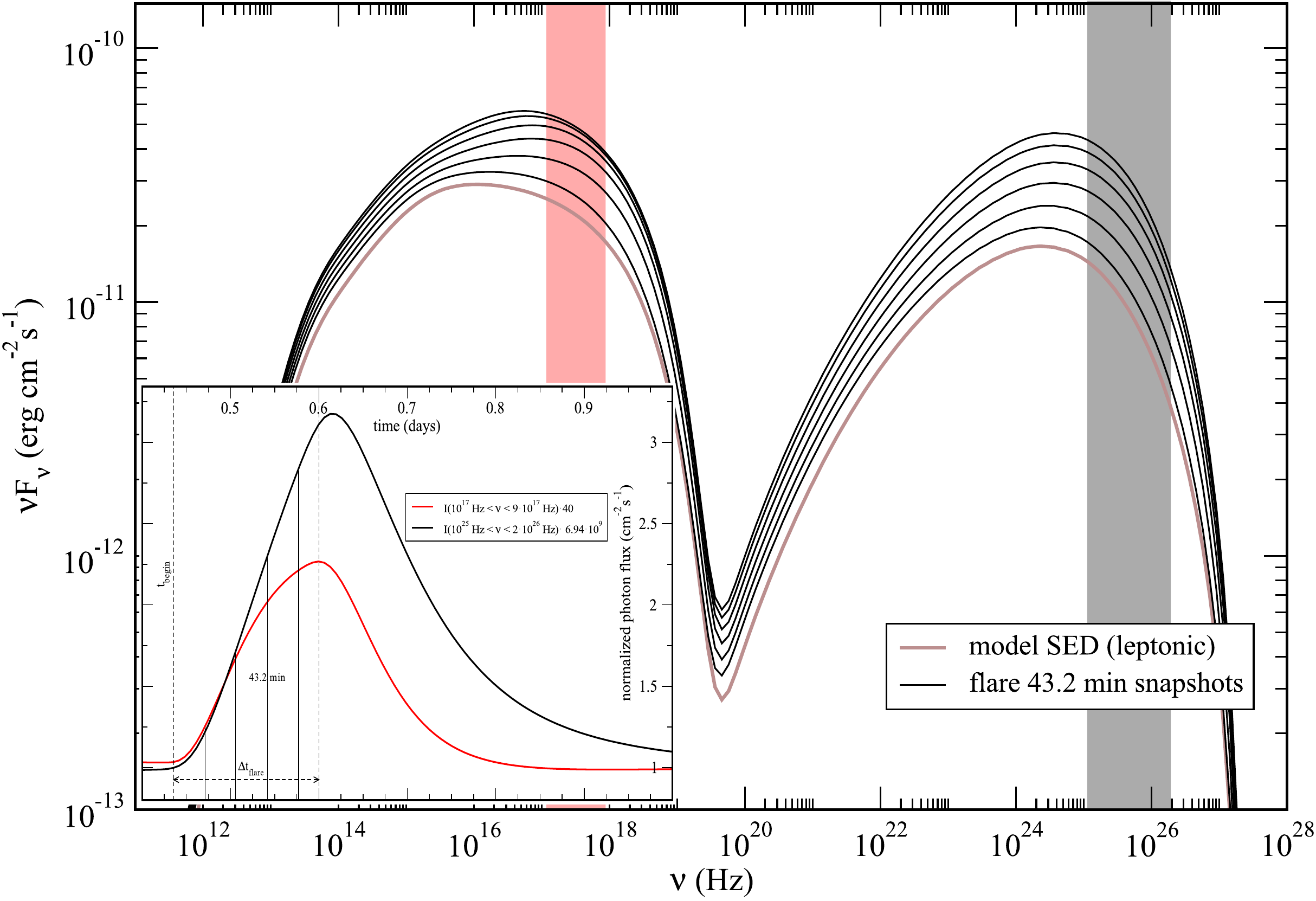}
\caption{Spectral evolution ($43.2$ min snapshots) during the imposed flaring scenario by setting $x_e = 2.5$ in the leptonic case of 1 ES 1011+496 (intrinsic spectrum). The SSC limit shows the simple flaring behaviour as expected, without spectral evolution in the VHE and orphan flares do not occur. The inset shows the corresponding lightcurves in the X-rays and gamma rays exhibiting a slight delay caused by the IC upscattering of the enhanced synchrotron photons.\label{fig7}}
\end{center}
\end{figure}
A flare occurs almost instantaneously in the optical ($2 \cdot 10^{14}$ Hz to $9 \cdot 10^{14}$ Hz) and X-ray ($1 \cdot 10^{17}$ Hz to $9 \cdot 10^{17}$ Hz) energy ranges due to the higher primary electron density accelerated to high energies.
Typical timescales of this first outburst are the acceleration and synchrotron loss timescale of electrons, see inset in Fig. \ref{fig5}. Along with the flare in the optical comes an outburst at VHE ($1 \cdot 10^{25}$ Hz to $2 \cdot 10^{26}$ Hz) with a slightly different rising behaviour due to enhanced reprocessed radiation of the initial photo-hadronic interactions, since more target (synchrotron) photons are provided during the outburst.
While the electron synchrotron peak shows spectral behaviour similar to SSC modelling, the photon index in the VHE softens as the cascaded radiation dominates over the proton synchrotron emission during the first flare, see Fig. \ref{fig6}. After approximately $0.7$ days the emission initiated by the enhanced $Q_{0,e^-}$ has cooled down to its steady-state value. The additionally injected protons (alongside the electrons) need to be accelerated with $t_{acc,p}$ to high $\gamma$s before emitting synchrotron radiation efficiently.
This causes the delay $t_{delay}$ between the first, MWL flare and the second, orphan flare in the VHE due to p$^+$-synchrotron radiation with its typical timescales, yet another feature not expected in a purely leptonic model. In contrast to the first outburst the photon index in the VHE hardens (inset of Fig. \ref{fig6}). Both are initiated by one single enhanced injection into the numerically considered region. Hence the VHE exhibit both characteristic timescales, those of electrons and those of protons in the magnetic field of the blob, and it is insufficient to limit theoretical investigation to $p^+$-timescales only.
\\\\In Fig. \ref{fig7} a hypothetical outburst with $x_e = 2.5$ and \mbox{$\Delta t = 3.86$} h in terms of the purely leptonic model (Table \ref{tbl1}a) is displayed for comparison, the inset shows the corresponding lightcurves. The behaviour is quite straightforward with its possible soft- and hard lags occurring, in contrast to the relatively complex time-dependent behaviour in the hybrid model. There is no noteworthy spectral evolution in the VHE and orphan flares do not occur. The timescales as shown in the lightcurves mainly differ from those of Table \ref{tbl1}b) as a result of the significantly lower magnetic field and are consistent with the modelling of comparable blazars \citep{weidinger02}.
\section{Discussion}
We have presented and applied a fully self-consistent and time-dependent hybrid emission model for blazar jets, taking all relevant processes into account, namely acceleration and synchrotron emission of electrons and protons, inverse Compton scattering, and intrinsic photo-hadronic interactions as well as $\gamma \gamma$-pair production. The model includes the self-consistent SSC limit by setting $Q_{0,p^+} \rightarrow 0$, see e.g. \citet{weidinger01} for a detailed description. However magnetic fields of $\mathcal{O}(10$ G$)$ will confine relativistic protons within the emitting region with a typical size of $\mathcal{O}(10^{15}$ cm$)$ to $\mathcal{O}(10^{16}$ cm$)$ as inferred from variability and causality \citep{weidinger01, boettcher03}. The numerical approach allows for a detailed treatment of the individual processes (see Chapter 2), including all non-linearities arising from the coupling of the photons to the leptons via photo-meson production and the creation of $e^{\pm}$-pairs with different
timescales. Hence the model spectra and lightcurves in case of outbursts become far more complex than in SSC models, even if non-linear cooling is included \citep{weidinger02, zacharias01}. Not only do orphan flares occur, but also variability associated with the different timescales for $p^+$ and $e^\pm$ in the system. More complex behaviour will arise if photon-quenching becomes effective, in the cases when the electromagnetic cascades serve as targets for photo-meson production. It is obvious that photo-hadronic cooling effects must not be neglected in those cases, but even without runaway production there is vast, qualitatively different, variable multiband behaviour, delicately depending on the chosen set of parameters, not only the behaviour 1 ES 1011+496 exhibits in our example. In addition typical time-lags are directly bound to the timescales, e.g. the acceleration timescale, in the system. Hence these appealing features could be used to determine typical acceleration rates in
astrophysical jets, otherwise inaccessible.
With future experiments a systematic approach of multiband variable behaviour is at hand, leading towards an investigation of the energy dependence of the involved timescales comparing lags and durations in different energy bands in our model, leading to advances in e.g. relativistic acceleration of particles, still not understood completely \citep{relshock01, relshock02, relshock03}.\\\\
1 ES 1011+496 was used as an example blazar with its low state emission already pointing out a relatively high magnetic field to be present within its jet, see Chapter 3. The steady state observed in the MWL campaign including the MAGIC telescopes and the Swift satellite is reliably described with our hybrid model, while a consistent leptonic approach fails to model the synchrotron peak of 1 ES 1011+496 (see Table 1 for the parameters used). One can of course assume various breaks with corresponding spectral indices in the electron distribution to describe the X-ray peak leptonically, but in general a physical motivation for these is rather hard to find. Hence, self-consistently speaking, the VHE peak is not due to inverse Compton photons, but consists of proton synchrotron radiation and emerging cascade radiation. We note that merely two additional parameters need to be introduced in the complex lepto-hadronic model with respect to the \citet{weidinger01} or \citet{tavecchio01} SSC models. The properties of the underlying electron and proton densities, like
the gap between $\gamma_{max,e}$ and $\gamma_{max,p}$ or spectral indices and breaks, arise consistently as a result of acceleration and cooling during the modelling process.\\
This has a certain impact on the variability observable from 1 ES 1011+496, investigated using an imposed flaring scenario simulating density fluctuations along the jet-axis (Eq. \ref{flare}). Quite obviously the timescales connected with $e^-$-synchrotron emission is lower than in the typical SSC-case where the magnetic fields are low. More importantly, one can observe patterns in the multiband lightcurves, see Fig. \ref{fig5}. In the imposed scenario, the first flare of 1 ES 1011+496 covers all considered energy bands with the timescale of the electrons, even in the VHEs, since it is caused by the freshly injected $e^-$ and the enhanced cascade radiation. With a time-lag of approximately $4.28$ days a second, orphan flare occurs in the VHEs because of $p^+$-synchrotron radiation with its typical timescale. It took the protons (with their higher $t_{acc,p}$) this time to be accelerated to the energies where they can emit efficiently. The fact that leptonic timescales are involved in the VHE emission as well, allows 
short-time variability in this region even in the hybrid case. This is not necessarily true in every hadronically dominated blazar, see e.g. \citet{boettcher02} where $p^+$-synchrotron radiation dominates over the cascades. In comparison with the outburst in the purely leptonic case, one can see that strong spectral features in the VHE during the flare are only present within the hybrid model with significant hardening and softening.\\\\
The multiband lightcurves can therefore be used as a hadronic fingerprint of an individual blazar by identifying the characteristic patterns using the presented model. The tangible patterns will be different in each blazar according to the physical parameters present within the jets, see e.g. \citet{weidinger03}, but typical timescales and timelags can only be found in the hybrid case and thus be used to identify blazars accelerating both electrons and hadrons without relying on the detection of neutrinos of an individual source, e.g. with IceCube, which is quite difficult \citep{neutrino00, neutrino01, neutrino02, neutrino03}. Consequently this paper suggests more long-term MWL observations of individual sources as done with Fermi LAT and VERITAS of Mkn 501 or Mkn 421 \citep{veritas, veritas02} to reveal those responsible for the recently detected extragalactic high energy neutrinos \citep{icecube_new}. Of course, AGN accelerating protons would be good candidates of ultra high energy cosmic ray production
sites among blazars, since proton energies up to $E_p = 10^{21}$ eV are reached in the observer's frame, and it
is thus important to identify the subset of those actually confining non-thermal protons, a task which can be addressed by our model. Systematic modelling of all types of blazars with one single model could also help to understand the differences and commonalities of each blazar flavour (FSQR, IBL, LBL, and HBL), because the parameters are obtained during the modelling. They are not biased using a specific Ansatz like leptonic or hadronic in the first place when producing the model spectrum. Thus, models like ours are crucial tools that can be used to interpret data in times of extensive multi-messenger campaigns and stacking astrophysics with the upcoming CTA, IceCube detecting first PeV neutrinos, and cosmic ray detectors like AUGER and HAWC in the future.

\begin{acknowledgements}
      We thank the referee for a helpful and detailed report.
      MW wants to thank Elitenetzwerk Bayern and the Graduate School 1147 for their support,
      FS acknowledged support from National Research Foundation through a ``Multiwavelength'' grant
\end{acknowledgements}


\end{document}